\documentclass[aps,pra,amssymb,twocolumn,superscriptaddress,10pt,showpacs]{revtex4}
\usepackage{graphicx}
\usepackage{dcolumn}
\usepackage{bm}
\usepackage{color}
\usepackage{subfigure}

\begin{document}

\newcommand{\beq}{\begin{equation}}
\newcommand{\eeq}{\end{equation}}
\newcommand{\barr}{\begin{eqnarray}}
\newcommand{\earr}{\end{eqnarray}}

\def\bra#1{\langle{#1}|}
\def\ket#1{|{#1}\rangle}
\def\sinc{\mathop{\text{sinc}}\nolimits}
\def\cV{\mathcal{V}}
\def\cH{\mathcal{H}}
\def\cT{\mathcal{T}}
\renewcommand{\Re}{\mathop{\text{Re}}\nolimits}
\newcommand{\tr}{\mathop{\text{Tr}}\nolimits}

\definecolor{dgreen}{rgb}{0,0.5,0}
\newcommand{\green}{\color{dgreen}}
\newcommand{\RED}[1]{{\color{red}#1}}
\newcommand{\BLUE}[1]{{\color{blue}#1}}
\newcommand{\GREEN}[1]{{\color{dgreen}#1}}
\newcommand{\REV}[1]{{\color{red}[[#1]]}}
\newcommand{\KY}[1]{\textbf{\color{red}[[#1]]}}
\newcommand{\SP}[1]{{\color{blue} [[#1. Saverio]]}}
\newcommand{\rev}[1]{{\color{red}[[#1]]}}

\def\HN#1{{\color{magenta}#1}}
\def\DEL#1{{\color{yellow}#1}}

\title{Typical observables in a two-mode Bose system}

\author{Paolo Facchi}
\affiliation{Dipartimento di Fisica and MECENAS, Universit\`a di Bari, I-70126 Bari, Italy}
\affiliation{INFN, Sezione di Bari, I-70126 Bari, Italy}

\author{Saverio Pascazio}
\affiliation{Dipartimento di Fisica and MECENAS, Universit\`a di Bari, I-70126 Bari, Italy}
\affiliation{INFN, Sezione di Bari, I-70126 Bari, Italy}

\author{Francesco V. Pepe}
\affiliation{Dipartimento di Fisica and MECENAS, Universit\`a di Bari, I-70126 Bari, Italy}
\affiliation{INFN, Sezione di Bari, I-70126 Bari, Italy}

\author{Golam Ali Sekh}
\affiliation{INFN, Sezione di Bari, I-70126 Bari, Italy}

\begin{abstract}
A class of $k$-particle observables in a two-mode system of Bose particles is
characterized by typicality: if the state of the system is sampled
out of a suitable ensemble, an experimental measurement of that
observable yields (almost) always the same result. We investigate
the general features of typical observables, the criteria to
determine typicality and finally focus on the case of density
correlation functions, which are related to spatial distribution of
particles and interference.
\end{abstract}


\maketitle

\section{Introduction}
\label{sec-intro}

The experimental realization of Bose-Einstein condensates raised
great theoretical interest. A system of Bose particles in one or
few single-particle states (modes) is an important workbench for
fundamental concepts in quantum mechanics and statistical physics
\cite{PW,SB,Leggett,BDZ,PS,PeSm,Leggettbook}. A large number of
particles distributed among two different modes, for example,
enables one to perform a full quantum double-slit experiment in a
single experimental run \cite{exptBEC}. Moreover, it has been
proposed that fluctuations in the interference patterns can probe
interesting characteristics of many-body systems
\cite{ref:NoiseCorr-AltmanDemlerLukin,ref:PolkovnikovAltmanDemlerPNAS,ref:GritsevAltmanDemlerPolkovnikov-NaturePhys2,ref:PolkovnikovEPL,ref:ImambekovGritsevDemlerVarenna,ref:GritsevDemlerPolkovnikov-PRA,ref:Hadzibabic-BECIntArray,ref:Hadzibabic-BECIntPhaseDefects,ref:Hadzibabic-BKT-Nature,ref:Hadzibabic2DTc,Raz}.

Interference is an interesting example of a property that
\textit{weakly} depends on the choice of the state of the system.
In a two-mode system, second-order-interference properties are
similar as far as one considers a number state or a phase state
\cite{CD,PS,PeSm}, while first-order properties are very
different. These features explain why an interference pattern can
be experimentally observed in single experimental runs by
measuring the particles' positions
\cite{JY,CGNZ,WCW,CD,PW,Leggett,BDZ}, although in the case of
number states the offset of the pattern fluctuates randomly, so
that averaging over a few experimental runs yields a flat density
profile. It is also interesting to note that, when the number of
particles is large, there are interference-related observables
whose experimental measurement yields the same result at each run
with overwhelming probability~\cite{AYI,YI,typbec}.

These considerations lead us to a general definition of
\textit{typicality} of an observable. Typicality is a mathematical
concept related to the phenomenon of measure concentration
\cite{Ledoux}. It has been used in many emerging phenomena in
physics and other sciences, with interesting applications on the
structure of entanglement in large quantum systems
\cite{Winter,FMPPS,mixmatrix,matrix reloaded,polarized}, and the
search for a quantum mechanical justification of some primary
statistical mechanical concepts
\cite{Lloyd,Tasaki,Gemmer,Popescu1,Popescu2,Goldstein,Sugita,Reimann,ref:Rigol-Nature,Cho,SugiuraShimizu}.
In this article, we shall apply the notion of typicality to a
two-mode Bose system with a fixed number of particles. We will
study the properties of observables with respect to uniform
sampling of a suitable Hilbert subspace. An observable will be
defined typical if each experimental run, performed on
\textit{any} state of the subspace, would yield the same result
with overwhelming probability. According to the probabilistic
interpretation of quantum mechanics, the expectation value of an
observable provides information on the average of experimental
runs on the (pure or mixed) state. We shall build on the results
of Ref.\ \cite{typbec} and prove that if the observable is
typical, there exists an expectation value that contains
information on (almost) each single experimental run, rather than
on the average result.

This article is organized as follows. In Section
\ref{sec-typicality} we specify the properties of the statistical
ensemble and introduce the general formalism, which leads to the
definition of typicality of an observable. In Section
\ref{sec:fluctgen} an analysis of the structure of fluctuations
for a general $k$-particle observable is performed. Quantitative
criteria will be given to determine typicality and obtain
informations on higher-order fluctuations. Section
\ref{sec:correlations} includes an application of typicality
criteria to density correlation operators, with the case study of
counterpropagating plane-wave modes and an extension to expanding
modes in the far-field regime.

\section{Typicality of an observable}
\label{sec-typicality}

Let us consider a system of bosons, without internal degrees of
freedom. The system can be described in a second-quantization
picture by introducing the annihilation and creation field
operators $\hat{\Psi}(\bm{r})$ and
${\hat{\Psi}}^{\dagger}(\bm{r})$, satisfying canonical equal-time
commutation relations:
\begin{eqnarray}
[\hat{\Psi}(\bm{r}), \hat{\Psi}(\bm{r}')] & = & 0, \\
\, [\hat{\Psi}(\bm{r}),  \hat{\Psi}^{\dagger}(\bm{r}')] & = &
\delta(\bm{r}-\bm{r}').
\end{eqnarray}
Once an orthonormal basis for the single-particle Hilbert space
has been chosen, the field operators can be expanded as sums of
mode operators, which create or annihilate a particle in one of
the basis states. We are going to analyze a system made up of $N$
bosons, distributed among two orthogonal modes, $a$ and $b$, with
single-particle wave functions $\psi_a(\bm{r})$ and
$\psi_b(\bm{r})$. A useful basis for the description of the
$N$-particle Hilbert space is formed by the Fock states
\begin{equation}\label{fock}
\ket{ \ell } := \left| \left( \frac{N}{2}+\ell
\right)_a, \left( \frac{N}{2}-\ell \right)_b \right\rangle,
\end{equation}
in which the two modes have well-defined occupation numbers. We
are assuming that $N$ is even for simplicity, but this
specification is immaterial in the large-$N$ limit. In a
second-quantized formalism, Fock states are obtained by applying a
sequence of mode creation operators to the vacuum $\ket{\Omega}$:
\begin{equation}
\ket{\ell} = \frac{1}{\sqrt{(N/2+\ell)! (N/2-\ell)!}}
({\hat{a}}^{\dagger})^{N/2+\ell}
({\hat{b}}^{\dagger})^{N/2-\ell} \ket{\Omega} .
\end{equation}
Since $\psi_a(\bm{r})$ and $\psi_b(\bm{r})$ are orthonormal, the
mode operators
\begin{equation}\label{modeop}
\hat{a} = \int d\bm{r}\, \psi_a^*(\bm{r}) \hat{\Psi}(\bm{r}), \qquad
\hat{b} = \int d\bm{r}\, \psi_b^*(\bm{r}) \hat{\Psi}(\bm{r})
\end{equation}
satisfy the canonical commutation relations
\begin{equation}
[ \hat{a}, \hat{a}^{\dagger} ] =  [ \hat{b},
\hat{b}^{\dagger} ] = 1,
\end{equation}
and all the operators of mode $a$ commute with those of mode $b$.
The number operators $\hat{N}_a={\hat{a}}^{\dagger} \hat{a}$ and
$\hat{N}_b={\hat{b}}^{\dagger} \hat{b}$ count the numbers of
particles in each mode.

We are interested in the properties of an ensemble of states
randomly sampled from the $n$-dimensional subspace
\begin{equation}
\mathcal{H}_n=\mathrm{span} \{ \ket{\ell} , |\ell| < n/2\},
\label{6}
\end{equation}
spanned by the Fock states with $0<n\leq N+1$ ($n$ is odd for
simplicity.) The case $n=1$ represents a degenerate ensemble in
which only the state $\ket{\ell=0}$ has nonvanishing probability
\cite{JY,CGNZ,WCW,CD,ref:PolkovnikovEPL,Paraoanu,YI}. We shall consider the large-$n$ case \cite{typbec}.
If a BEC made up on $N$ particles is ``evenly" split between the two modes, one expects $n = O(\sqrt{N})$. However, more general cases are possible \cite{typbec}. In the following we will assume that
\begin{equation}
n = o(N).
\end{equation}
The assumption of \emph{uniform} sampling is clearly a simplifying
one: the number of states that are involved in the description and
their amplitudes depends on the experimental procedure, yielding
the separation of the condensate in the two modes \cite{esteve}.
However, it will emerge that our main results are qualitatively
unchanged for a large and relevant class of probability
distributions on $\mathcal{H}_n$.

The average of the  projection on a random vector state $\ket{\Phi_N} \in \mathcal{H}_n$ uniformly sampled on $\mathcal{H}_n$ yields the (microcanonical) density matrix
\begin{eqnarray}
\hat{\rho}_n =  \overline{\ket{\Phi_N}\bra{\Phi_N}}
= \frac{1}{n}
\sum_{|\ell| < n/2 } \ket{\ell} \bra{\ell} =: \frac{1}{n}
\hat{P}_n ,
\label{rhon}
\end{eqnarray}
where $\hat{P}_n$ is the projection onto the subspace
$\mathcal{H}_n$. In the following, only the density matrix
(\ref{rhon}) will explicitly appear in our analysis and
calculation. The definition and use of the random state
$\ket{\Phi_N}$ in Eq.~(\ref{rhon}) is superfluous and can be
dispensed with, in accord with the prescription of Ockham's razor.
In this respect, it is worth stressing that no hypothesis of
decoherence will be made and if one wants one can safely assume
that in each experimental run a wave function describes the
condensate, which is therefore in a pure state.

Given an observable $\hat{A}$, the statistical average of its
expectation value reads
\begin{eqnarray}
\label{expGEN}
\overline{A}
 :=  \tr (\hat{\rho}_n \hat{A} ) = \frac{1}{n} \sum_{|\ell| <
n/2 } \bra{\ell} \hat{A} \ket{\ell}.
\end{eqnarray}
Its quantum variance is
\begin{equation}\label{deltahatA}
\delta A^2 := \tr (\hat{\rho}_n \hat{A}^2 ) - {\!\left[ \tr
(\hat{\rho}_n \hat{A} ) \right]\!}^2.
\end{equation}
The significance of this quantity in the context of ensemble
statistics becomes clear once it is decomposed in the sum of two contributions \cite{FNPPSY,typbec}: it contains both the classical and quantum uncertainties of the  observable $A$ in the microcanonical  state $\rho_n$.
Therefore, the asymptotic condition
\begin{equation}
\delta A= o(\overline{A}),
\end{equation}
for $N\to\infty$, ensures that in the overwhelming majority of cases the
experimental measurement of the observable $\hat{A}$ will
fluctuate within an extremely narrow range around the average
expectation value $\overline{A}$. Thus, the outcome of a
measurement of $\hat{A}$ is almost always the same (and it equals its average) for every experimental run  in the ensemble. We call this property \textit{typicality of the observable}.

In the following sections we will
define and characterize conditions for a $k$-particle observable
to be typical, and analyze in detail the case of spatial
correlation functions, which are related to interference.

\section{Control of fluctuations for a $k$-particle operator}
\label{sec:fluctgen}

In a second quantization formalism, the field operators $\hat{\Psi}$ and $\hat{\Psi}^{\dagger}$ can be used to build up
many-body observables \cite{fetter}. The simplest ones are the
Hermitian $2k$-point functions
\begin{equation}
\label{Gk}
\hat{G}_k (\bm{r}_1,\dots,\bm{r}_k) :=
\prod_{i=1}^k \hat{\Psi}^{\dagger}(\bm{r}_i) \hat{\Psi}(\bm{r}_i).
\end{equation}
Following Eq.\ (\ref{expGEN}), one can observe that the ensemble
average of $\hat{G}_k$ can be computed for any $n$ once its expectation value over the $N$-particle-two-mode Fock states $\ket{\ell}$ is known. Thus, when the field operators in (\ref{Gk}) are expanded in orthogonal modes, only the terms formed by operators associated to modes $a$ and $b$ are relevant in $\overline{G_k}$.
Moreover, operator products with a different number of
$\hat{a}^{\dagger}$'s and $\hat{a}$'s (respectively $\hat{b}^{\dagger}$'s and
$\hat{b}$'s) give vanishing contributions to $\bra{\ell} \hat{G}_k
\ket{\ell}$. The relevant terms in (\ref{Gk}) can be reordered to
be expressed as number operators $\hat{N}_{a,b}$. The average
eventually reads [we will assume $k=O(1)$]
\begin{eqnarray}\label{avGk}
& & \overline{ G_k (\bm{r}_1,\dots,\bm{r}_k) } = \sum_{m=0}^k F_m
(\bm{r}_1,\dots,\bm{r}_k) \nonumber \\ & & \times \frac{1}{n}
\sum_{|\ell| <n/2} \prod_{A=0}^{m-1}
\!\left(\frac{N}{2}+\ell-A\right)\! \prod_{B=0}^{k-m-1}
\!\left(\frac{N}{2}-\ell-B\right)\! , \nonumber \\
\end{eqnarray}
with
\begin{equation}\label{Fm}
F_m (\bm{r}_1,\dots,\bm{r}_k) = \left| \Phi_m
(\bm{r}_1,\dots,\bm{r}_k) \right|^2,
\end{equation}
where $\Phi_m (\bm{r}_1,\dots,\bm{r}_k)$ is, apart from a normalization factor, the symmetrized $k$-body wave function with
$m$ particles in mode $a$ and $k-m$ particles in mode $b$:
\begin{eqnarray}
\label{Phim}
& & \Phi_m (\bm{r}_1,\dots,\bm{r}_k) = \nonumber \\
& & \sum_{\sigma}
\psi_a(\bm{r}_{\sigma(1)})\dots\psi_a(\bm{r}_{\sigma(m)})
\psi_b(\bm{r}_{\sigma(m+1)}) \dots \psi_b(\bm{r}_{\sigma(k)}),
\nonumber \\
\end{eqnarray}
$\sigma$ denoting permutation of $k$ elements.

An important class of $k$-particle observables  can be obtained by
integrating the $2j$-point functions, for $j\leq k$, with a
multiplicative kernel $\mathcal{A}_j$ \footnote{We shall focus on
this case for simplicity, but the formalism, as well as the
results, can be generalized to kernels involving derivatives. On
the other hand, momentum-dependent observables can be more easily
treated by using correlation functions in momentum space,
constructed with the Fourier-transformed field operators
$\hat{\tilde{\Psi}}(\bm{p}) = \int d\bm{r} e^{-i\bm{p}\cdot\bm{r}}
\hat{\Psi}(\bm{r})$.}:
\begin{equation}
\label{Ak}
\hat{A}_k = \sum_{j=0}^k \int d\bm{r}_1 \dots d\bm{r}_j
\mathcal{A}_j (\bm{r}_1,\dots\bm{r}_j) \hat{G}_j
(\bm{r}_1,\dots\bm{r}_j).
\end{equation}
The ensemble average of $\hat{A}$ can be immediately computed
inserting the general results (\ref{avGk}). Thus, since
$\overline{G_j}=O(N^j)$ when $k=O(1)$, we have
\begin{equation}
\overline{A_k} = \int d\bm{r}_1 \dots d\bm{r}_k \mathcal{A}_k
(\bm{r}_1,\dots\bm{r}_k) \overline{G_k (\bm{r}_1,\dots\bm{r}_j)} +
O(N^{k-1}).
\end{equation}
In order to study the behavior of the variance (\ref{deltahatA})
and determine whether $\hat{A}_k$ is typical or not, we should
compute its square, which involves at the highest order in $N$ a
product of $\hat{G}_k$ functions, that can be recast into a single
$4k$-point function by normal-ordering the field operators:
\begin{eqnarray}\label{Aksquare}
\hat{A}_k^2 & = & \int d\bm{r}_1 \dots d\bm{r}_{2k} \bigl(
\mathcal{A}_k(\bm{r}_1,\dots,\bm{r}_k)
\mathcal{A}_k(\bm{r}_{k+1},\dots,\bm{r}_{2k}) \nonumber \\ & &
\times \hat{G}_{2k} (\bm{r}_1,\dots,\bm{r}_{2k}) \bigr) +
O(N^{2k-1}).
\end{eqnarray}
From~(\ref{Ak})-(\ref{Aksquare}) we get an expression for
the variance
\begin{eqnarray}
\label{deltaAk}
\delta A_k^2\!&\!=\!&\!\int\!d\bm{r}_1 \dots d\bm{r}_{2k}
\bigl( \mathcal{A}_k(\bm{r}_1,\dots,\bm{r}_k)
\mathcal{A}_k(\bm{r}_{k+1},\dots,\bm{r}_{2k}) \nonumber \\ & &
\times \gamma_{k} (\bm{r}_1,\dots,\bm{r}_{2k}) \bigr) +
O(N^{2k-1}),
\end{eqnarray}
with
\begin{eqnarray}
\gamma_k (\bm{r}_1,\dots,\bm{r}_{2k}) &:=& \overline{G_{2k}(\bm{r}_1,\dots,\bm{r}_{2k})}
\nonumber \\ &-&\overline{G_{k}(\bm{r}_1,\dots,\bm{r}_{k})} \;\;
\overline{G_{k}(\bm{r}_{k+1},\dots,\bm{r}_{2k})}, \nonumber \\
\label{gammak}
\end{eqnarray}
that depends only on the ensemble averages of $G_{k}$ and
$G_{2k}$. Notice that the order $N^{2k-1}$ comes both from
contributions in (\ref{Ak}) with $j<k$ and from normal ordering.

Our aim is to find whether, for some choice of the sampled Hilbert
subspace $\mathcal{H}_n$, the standard deviation (\ref{deltaAk})
is of smaller order with respect to the average $\overline{A_k}$ in the
asymptotic $N\to\infty$ regime. This behavior is in accord with the definition of typicality of the observable $\hat{A}_k$ given in Sec.\ \ref{sec-typicality}. First, let us observe that, since $\overline{G_k}$ is polynomial in $N$ and $n$, inserting the general result (\ref{avGk}) into $\delta A_k^2$ yields a polynomial function of degree $2k$ in the number of particles and
the dimension of the sampled subspace:
\begin{equation}\label{deltaAkpoly}
\delta A_k^2 = \sum_{p=0}^k \sum_{q=0}^{2k-p}
D_{p,q}^{({A}_k)} \!\left( \frac{N}{2} \right)^p n^q.
\end{equation}
Due to the symmetry of the summand around $\ell=0$, only terms
$N^{2k-q}n^q$ with {\it even} $q$ are present in
(\ref{deltaAkpoly}). It is evident that, as far as $n=o(N)$, it is
necessary and sufficient for $\hat{A}_k$ to be typical that
\begin{eqnarray}\label{typ1}
D_{2k,0}^{({A}_k)} & = & \int d\bm{r}_1 \dots d\bm{r}_{2k}
\mathcal{A}_k(\bm{r}_1,\dots,\bm{r}_k)
\mathcal{A}_k(\bm{r}_{k+1},\dots,\bm{r}_{2k}) \nonumber \\ & &
\times \biggl( \sum_{M=0}^k F_M (\bm{r}_1,\dots,\bm{r}_{2k}) -
\nonumber \\ & & \sum_{m,m'=0}^k F_{m'}
(\bm{r}_1,\dots,\bm{r}_{k}) F_m (\bm{r}_{k+1},\dots,\bm{r}_{2k})
\biggr)
\end{eqnarray}
vanish, namely,
\begin{equation}
\label{eq:typcond}
D_{2k,0}^{({A}_k)} = 0.
\end{equation}
 In this case, since $\overline{A_k}=O(N^k)$, the relative
fluctuations are
\begin{equation}\label{relfluc}
\frac{\delta A_k}{\overline{A_k}} = O \!\left(
\frac{1}{\sqrt{N}} \right) + O \!\left(\frac{n}{N} \right) \to 0 \qquad \text{ for }
N\to\infty.
\end{equation}
Fluctuations that scale like $N^{-1/2}$, related to normal ordering and to the very definition of $\hat{A}_k$, are ensemble-independent, in the sense that they are present even in degenerate distributions of states. Linear fluctuations in $n$ are clearly related to the dimension of the sampled subspace, and therefore strongly depend on the definition of the ensemble. It is interesting to note that, if $n=O(N^{\alpha})$, two qualitative regimes can be distinguished: i) when $\alpha\leq 1/2$, the relative fluctuations scale like $N^{-1/2}$; ii) while if $\alpha>1/2$, they asymptotically vanish like $N^{\alpha-1}$, i.e.\ more slowly. This reasoning is based on the assumption that
$D_{2k-2,2}^{({A}_k)}$ does not vanish: we will see in the
following how this condition can be checked and discuss in the
next section a relevant class of exceptions.

Due to the form of the symmetrized products of wave functions
(\ref{Phim}), the typicality condition~(\ref{eq:typcond})
can be recast in a more convenient form: in particular, we can
dispose of the integral over $2k$ variables by observing that the
functions $\Phi_M(\bm{r}_1,\dots,\bm{r}_{2k})$ can be
decomposed as
\begin{eqnarray}\label{clusters}
& & \Phi_M (\bm{r}_1,\dots,\bm{r}_{2k}) = \nonumber \\ & &
\sum_{m=\max \{ 0,M-k \}}^{\min \{ M,k \}} \Phi_{M-m}
(\bm{r}_1,\dots,\bm{r}_{k}) \Phi_m
(\bm{r}_{k+1},\dots,\bm{r}_{2k}). \nonumber \\
\end{eqnarray}
This result is related to the cluster decomposition principle in Bose systems~\cite{weinberg}. When one computes the square modulus $F_M=|\Phi_M|^2$ of (\ref{clusters}), that enters the variance in (\ref{gammak}), the term $\sum_m F_{M-m} F_m$ appears. This contribution exactly cancels with the subtracted terms in (\ref{gammak}), since it can be obtained by a change of summation indices
\begin{eqnarray}
\label{sums}
& & \sum_{m,m'=0}^{k} F_{m'} (\bm{r}_1,\dots,\bm{r}_{k}) F_m
(\bm{r}_{k+1},\dots,\bm{r}_{2k}) = \nonumber \\ & & \sum_{M=0}^{k}
\sum_{m=0}^{M} F_{M-m} (\bm{r}_1,\dots,\bm{r}_{k}) F_m
(\bm{r}_{k+1},\dots,\bm{r}_{2k}) \nonumber \\ & + &
\sum_{M=k+1}^{2k} \sum_{m=M-k}^{k} F_{M-m}
(\bm{r}_1,\dots,\bm{r}_{k}) F_m (\bm{r}_{k+1},\dots,\bm{r}_{2k}).
\nonumber \\
\end{eqnarray}
Accordingly, a compact form of the typicality condition~(\ref{eq:typcond}) reads
\begin{equation}\label{typ2}
D_{2k,0}^{({A}_k)} = \sum_{M=0}^{2k} \sum_{m'\neq m=\max \{ 0,M-k \}}^{\min \{ M,k \}} \mathcal{I}_{M-m,M-m'}^{({A}_k)}
\mathcal{I}_{m,m'}^{({A}_k)} = 0,
\end{equation}
where the coefficients $\mathcal{I}_{m,m'}$ are integrals over $k$ position variables of products of the type $\Phi_{m'}^*\Phi_m^{ }$:
\begin{equation}\label{intI}
\mathcal{I}_{m,m'}^{({A}_k)}\!=\!\int \! d\bm{r}_1\dots
d\bm{r}_k \mathcal{A}_k(\bm{r}_1,\dots,\bm{r}_k) \!\left(\Phi_{m'}^* \Phi_m^{ } \right) (\bm{r}_1,\dots,\bm{r}_k).
\end{equation}
Given the integral kernel $\mathcal{A}_k$, which determines the
highest order in $N$ of the observable $\hat{A}_k$, and the mode
wave functions $\psi_{a,b}(\bm{r})$, it is sufficient to compute
the integrals (\ref{intI}) to check whether the observable is
typical.

It is also interesting to analyze the structure of
$D_{2k-2,2}^{({A}_k)}$, also in view of the following
discussion on spatial correlation function. If
$D_{2k-2,2}^{({A}_k)}$ in nonvanishing, the transition between
an ensemble-independent $(N^{-1/2})$ and an ensemble-dependent
behavior of fluctuations occurs for $n=O(N^{1/2})$. In order to
analyze this quantity, one should take into account terms of order
$N^k$ and $N^{k-2}n^2$ in the general expression (\ref{avGk})
\begin{eqnarray}
& & \overline{ G_k (\bm{r}_1,\dots,\bm{r}_k) } = \sum_{m=0}^k F_m
(\bm{r}_1,\dots,\bm{r}_k) \nonumber \\ & & \times \!\left[
\!\left( \frac{N}{2} \right)^k + \!\left( \frac{N}{2}
\right)^{k-2} \frac{n^2}{24} (k^2-k(4m+1)+4m^2) \right] \nonumber \\
& & + O(N^{k-1})+ O(N^{k-4}n^4),   \label{Gkbar}
\end{eqnarray}
to be used in the computation of (\ref{gammak}). Integration over
the kernel $\mathcal{A}_k$ and application of the same change of
indices leading to Eq.\ (\ref{sums}) yield the result
\begin{eqnarray}\label{Dn2}
&\!& D_{2k-2,2}^{({A}_k)} = \frac{1}{12} \Bigl\{ \bigl(
\mathcal{J}^{({A}_k)} \bigr)^2 \nonumber \\ &\!& +
\sum_{M=0}^{2k} \sum_{m'\neq m=\max \{ 0,M-k \}}^{\min \{ M,k \}}
\bigl[ \mathcal{I}_{M-m,M-m'}^{({A}_k)}
\mathcal{I}_{m,m'}^{({A}_k)} \nonumber \\
&\!& \times (2k^2-k(4M+1)+2M^2) \bigr] \Bigr\},
\end{eqnarray}
where the $\mathcal{I}$ integrals have been defined in
(\ref{intI}), and
\begin{eqnarray}\label{intJ}
\mathcal{J}^{({A}_k)} & = & \int d\bm{r}_1\dots d\bm{r}_k
\mathcal{A}_k (\bm{r}_1,\dots,\bm{r}_k) \nonumber \\ & & \times
\sum_{m=0}^{k} F_m (\bm{r}_1,\dots,\bm{r}_k) (k-2m).
\end{eqnarray}
The summation appearing in (\ref{intJ}) contains an equal number
of positive and negative term. Thus, it can vanish if the mode
structure is properly chosen. In particular, it vanishes when the $F_m$'s are invariant with respect to exchange of the two mode
wave functions:
\begin{equation}
\label{Fmsymm}
F_m = F_{k-m} \quad \Rightarrow \quad \sum_{m=0}^k F_m (k-2m)=0.
\end{equation}
This condition is always valid in a ``double-slit" BEC
interference experiment
\cite{exptBEC,Schmied,Schmied2,Schmied3,Schmied4,Schmied5}, where
the two modes are identically prepared (within experimental
accuracy) and then let to interfere. It is easy to check that the
above condition is satisfied whenever
$|\psi_a(\bm{r})|=|\psi_b(\bm{r})|$ at all points. In turn, this
is a consequence of the invariance of $F_m$ under local phase
transformations $\psi_{a,b}(\bm{r})\to
e^{i\varphi(r)}\psi_{a,b}(\bm{r})$.

\section{Typicality of density correlations}
\label{sec:correlations}

In this section we will specialize the general results obtained for
an observable of the form (\ref{Ak}) to a particular class, related to the spatial interference of condensates. This is usually
accessible to experimentalists and provides an interesting example of typical behavior. We shall analyze the integrated
density correlation functions
\begin{eqnarray}
\label{Ck}
\hat{C}_k(\bm{x}_1,\dots,\bm{x}_{k-1}) & = & \int d\bm{r}
\hat{\rho}(\bm{r}) \hat{\rho}(\bm{r}+\bm{x}_1)\dots
\hat{\rho}(\bm{r}+\bm{x}_{k-1}) \nonumber \\ & = & \int d\bm{r}
\hat{G}_k (\bm{r},\bm{r}+\bm{x}_1,\dots,\bm{r}+\bm{x}_{k-1}) \nonumber \\
& & + O(N^{k-1}),
\end{eqnarray}
where $\hat{\rho}(\bm{r})=\hat{G}_1(\bm{r})$.
In agreement with Eq.\ (\ref{Ak}), it is clear from the normal-ordered form that the highest-order integral kernel for
this class of observables is
\begin{equation}
\label{Ckernel}
\mathcal{C}_k (\bm{r}_1,\dots,\bm{r}_k) = \prod_{i=1}^{k-1}
\delta(\bm{r}_{i+1}-\bm{r}_i-\bm{x}_i).
\end{equation}
The singularities arising in the normal ordering of (\ref{Ck}) can be avoided by smearing all densities around the points $(\bm{r}+\bm{x}_i)$ with functions that take into account the finite experimental spatial resolution. We will focus on two modes that are paradigmatic in the description of
interference of Bose-Eisntein condensates, namely two
counterpropagating plane waves
\begin{equation}\label{waves}
\psi_a(\bm{r}) = e^{i \bm{k}_0 \cdot \bm{r}}, \qquad
\psi_b(\bm{r}) = e^{-i \bm{k}_0 \cdot \bm{r}},
\end{equation}
and eventually extend the results to the case of two modes that are spatially separated at the initial time, and are let to expand and overlap: this is a more realistic description of the experiments.

\begin{figure}
\centering
\includegraphics[width=0.45\textwidth]{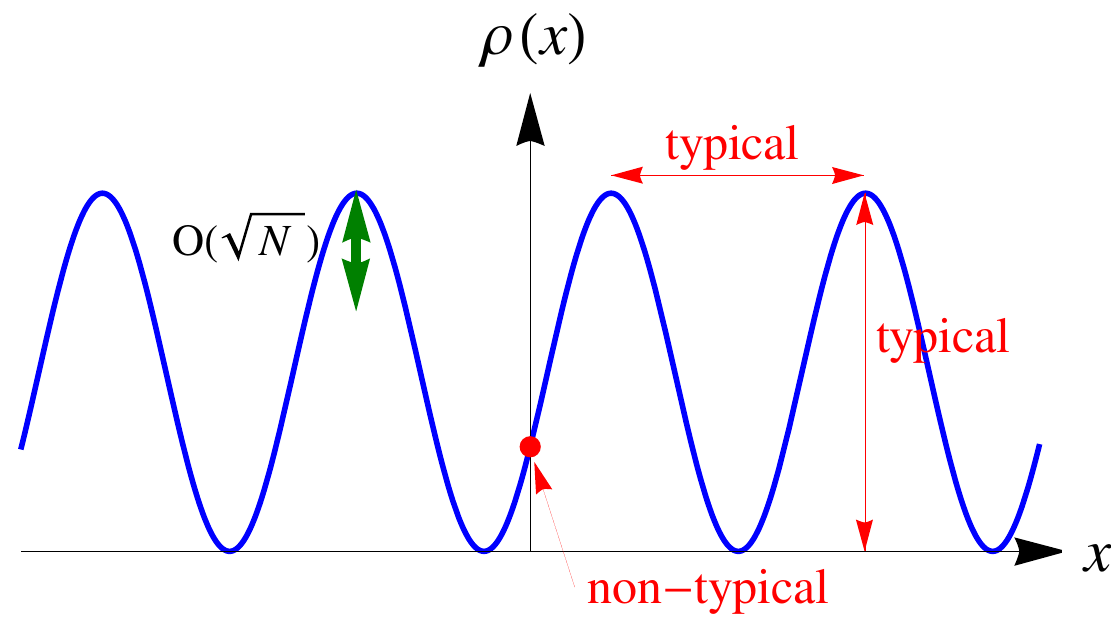}
\caption{Outcome of a density measurement in a two plane-wave mode
system, in the asymptotic regime. The visibility and the period of
the interference pattern are (almost) fixed by typicality, while
the offset of the pattern, and thus the value of density at the
origin, fluctuates randomly. The order $\sqrt{N}$ of the
visibility fluctuations refers to the case $n=O(N^{\alpha})$ with
$\alpha\leq 3/4$.}\label{fig:typ}
\end{figure}

In order to satisfy the typicality condition (\ref{typ2}), it is generally not necessary that the $\mathcal{I}$ integrals
themselves vanish. However, in this case we can verify the stronger condition that all the integrals with $m'\neq m$ are identically zero. Let us compute the general form of these integrals by using
(\ref{Ckernel})
\begin{eqnarray}\label{intIC}
\mathcal{I}_{m',m}^{({C}_k)} &\!=\!&\!\int\!d\bm{r}_1\dots
d\bm{r}_k \mathcal{C}_k (\bm{r}_1,\dots,\bm{r}_k) \!\left(
\Phi_{m'}^* \Phi_m^{ } \right)\! (\bm{r}_1,\dots,\bm{r}_k)
\nonumber
\\ &\!=\!&\!\int\!d\bm{r} \!\left(
\Phi_{m'}^* \Phi_m^{ } \right)\!
(\bm{r},\bm{r}+\bm{x}_1,\dots,\bm{r}+\bm{x}_{k-1}),
\end{eqnarray}
and consider for definiteness $m'>m$. In this case, the function
$\Phi_{m'}^*\Phi_m^{ }(\bm{r}_1,\dots,\bm{r}_k)$ is the sum of products of $k$ mode wave functions and $k$ complex conjugates, with the number of $\psi_a^*$'s exceeding the number of $\psi_a$'s by$m'-m$. The structure of the products reads
\begin{eqnarray}\label{prodC}
& & \prod_{\sigma=1}^S |\psi_a(\bm{r}_{j_{\sigma}})|^2
\prod_{\tau=1}^T |\psi_b(\bm{r}_{j_{\tau}})|^2 \nonumber \\ &
& \times \prod_{\zeta=1}^Z \bigl(\psi_a^*\psi_b^{ } \bigr)
(\bm{r}_{j_{\zeta}}) \prod_{\xi=1}^X \bigl(\psi_b^*\psi_a^{ }
\bigr) (\bm{r}_{j_{\xi}})
\end{eqnarray}
with $S+T+Z+X=k$, $S+Z=m'$ and $S+X=m$, which implies
\begin{equation}
Z-X=m'-m
\end{equation}
for all products in $\Phi_{m'}^*\Phi_m^{ }$. For the plane-wave
modes (\ref{waves}), $|\psi_{a,b}|=1$ and $\psi_a^*\psi_b^{ }=
e^{-2i\bm{k}_0\cdot\bm{r}}$. Inserting these results in the
general form of the products (\ref{prodC}) and integrating over
$\bm{r}$ as in (\ref{intIC}) yields
\begin{equation}
\prod_{\zeta=1}^Z e^{-2i\bm{k}_0\cdot\bm{x}_{j_{\zeta}-1}}
\prod_{\xi=1}^X e^{2i\bm{k}_0\cdot\bm{x}_{j_{\xi}-1}} \int
d\bm{r} e^{-2i(m'-m)\bm{k}_0\cdot\bm{r}} = 0.
\end{equation}
Thus, all the contributions to the integral (\ref{intIC})
identically vanish for every $k$ and every set of points
$(\bm{x}_1,\dots,\bm{x}_{k-1})$. This implies that condition
\begin{equation}\label{wtyp1}
D_{2k,0}^{({C}_k)} = 0
\end{equation}
is satisfied by all density correlation functions of the form
(\ref{Ck}), which are thus typical for $n=o(N)$. The structure of the modes provides interesting information also on the
$N^{2k-2}n^2$ part of the fluctuations, arising from (\ref{gammak}) and the general expression (\ref{Gkbar}). Since $|\psi_a|$ and $|\psi_b|$ are identically equal to one, the mode wave functions
satisfy the symmetry condition (\ref{Fmsymm}) for the $F_m$ functions.
This implies that the $\mathcal{J}^{({C}_k)}$ integrals,
defined as in Eq.~(\ref{intJ}), indentically vanish, which,
together with the cancellation of the $\mathcal{I}$ integrals,
leads to the result
\begin{equation}\label{wtyp2}
D_{2k-2,2}^{({C}_k)} = 0.
\end{equation}
Thus, the highest order of ensemble-dependent contributions to the
variance $\delta C_k^2$ is indeed $N^{2k-4}n^4$, which
should be compared with the order $N^{2k-1}$ of the ensemble-independent fluctuations. This means that if $n=O(N^{\alpha})$, the relative fluctuations behave like
\begin{equation}
\delta C_k^2 = \left\{ \begin{array}{ll} O\left(
\frac{1}{\sqrt{N}} \right) & \text{ for } \alpha\leq 3/4, \\ & \\
O\left( \left(\frac{n}{N}\right)^4 \right)= O(N^{4(\alpha-1)}) &
\text{ for } \alpha> 3/4.
\end{array} \right.
\end{equation}
The transition between an ensemble-independent regime of relative fluctuations to an ensemble-dependent one does not take place, as one would expect, at $n \sim \sqrt{N}$, but extends up to $n \sim N^{3/4}$. This behavior, which is general for density correlation functions, generalizes the results obtained in~\cite{typbec} for
the Fourier transform of the second order density correlation function
\begin{equation}\label{C2}
\hat{C}_2 (\bm{x}) = \int d\bm{r} \hat{\rho}(\bm{r})
\hat{\rho}(\bm{r}+\bm{x}).
\end{equation}
We include a brief discussion on this operator to clarify what
implications typicality has from an experimental point of
view. The expectation value of (\ref{C2}) is generally given by
\begin{equation}\label{avC2}
\overline{C_2(\bm{x})} = \frac{N^2}{4} \sum_{m=0}^2 \int d\bm{r}
F_m(\bm{r},\bm{r}+\bm{x}) + O(N)+O(n^2),
\end{equation}
which, in the case of plane waves, specializes to
\begin{equation}\label{avC2w}
\overline{C_2(\bm{x})} \simeq N^2 \left[ 1 + \frac{1}{2} \cos
(2\bm{k}_0\cdot\bm{x}) \right].
\end{equation}
Due to typicality, the function (\ref{avC2w}) represents the overwhelmingly probable experimental result of a measurement of
the observable $\hat{C}_2(\bm{x})$, unless $n=O(N)$. This function
is the two-point density correlation of a {\it
classical} density (see discussion in Ref.~\cite{typbec})
\begin{equation}
\rho(x) = 2 N^2 \cos^2 (\bm{k}_0\cdot\bm{x} + \phi).
\end{equation}
Among all the parameters that determine the typical experimental
outcome of a density measurement, one, namely the offset $\phi$ of
the interference pattern, is not determined by typicality, since
correlation functions do not depend on it. In Figure \ref{fig:typ}
a possible outcome of a density measurement in a two plane-wave
mode system is represented, with its typical features highlighted.

Let us finally discuss the typicality properties of density
correlation functions for a two-mode system that is closer to
actual experimental implementations, and find out in which cases
the results (\ref{wtyp1})-(\ref{wtyp2}) can be generalized. Let us
consider two spatially separated modes $\psi_a(z)$ and $\psi_b(z)$
in one dimension, which are concentrated respectively around
positions $z_a$ and $z_b$, their initial width being much smaller
than their distance. The spatial separation implies that the
convolution
\begin{equation}\label{convol}
\int dz' \psi_a^*(z') \psi_b(z'+z) =:
e^{F_{ab}(z)+i\varphi_{ab}(z)},
\end{equation}
with $F_{ab}$ and $\varphi_{ab}$ real functions of $z$, is peaked
around a point $z=z_0$, while the convolutions of $\psi_a^*$ with
$\psi_a$ and of $\psi_b^*$ with $\psi_b$ are peaked around $z=0$.
If the trapping potential is turned off at the initial time $t=0$,
the particles evolve under the free Hamiltonian $H_0=-d^2/dz^2$.
In absence of collisions, the $k$-point functions at $t>0$ can be
obtained by replacing
\begin{equation}
\psi_{a,b}(z) \quad \to \quad \psi_{a,b}(z,t)=\exp\!\left( - i t
H_0 \right) \psi_{a,b}(z).
\end{equation}
We are interested in the large-time ({\it far-field}) regime, in
which the evolved wave function are approximated by the asymptotic
form \cite{teschl}
\begin{equation}\label{farfield}
\psi_{a,b} (z,t) \simeq \left( \frac{1}{4\pi i t}
\right)^{\frac{1}{2}} e^{\frac{i x^2}{4t}}
\tilde{\psi}_{a,b}\!\left( \frac{z}{2t} \right),
\end{equation}
where $\tilde{\psi}_{a,b}(k)=\int dz e^{-ikz} \psi_{a,b}(z)$ are
the Fourier transforms of the {\it initial} modes.
In the far-field approximation, the quantities $\psi_a^*\psi_b^{}$
and their complex conjugates, which enter the typicality condition
(\ref{typ2}) through products of the type (\ref{prodC}), are thus
related to products of Fourier transforms, which can be
expressed through the convolution (\ref{convol}) as
\begin{equation}
\tilde{\psi}_a^* (k) \tilde{\psi}_b^{ } (k) = \int dz \,
e^{F_{ab}(z)+i\varphi_{ab}(z)-ikz}.
\end{equation}
Using a quadratic approximation of $F_{ab}$ around its maximum
$z_0$, the integral becomes Gaussian and the product
$\psi_a^*\psi_b^{}$ reads
\begin{eqnarray}\label{psiab}
\psi_a^* (z,t) \psi_b^{} (z,t) & \simeq & c(z_0,t) \exp\!\left( -
i \frac{z_0 z}{2t} \right)\! \nonumber \\ & & \times \exp\!\left(
- \frac{z^2}{8 t^2 |F''_{a,b}(z_0)|} \right) ,
\end{eqnarray}
If the spatial period of the complex exponential is much shorter
than the standard deviation of the Gaussian part, namely
\begin{equation}\label{distance}
z_0 \gg \frac{2\pi}{|F''_{a,b}(z_0)|},
\end{equation}
the product (\ref{psiab}) is consistent with the result for
counterpropagating plane wave modes with time-dependent wave
number $k_0(t)=z_0/4t$, modulated by a slowly-varying Gaussian
envelope, whose width is of the same order as the standard
deviation of the far-field-approximated $|\psi_a|^2$ and
$|\psi_b|^2$. Thus, the integrals of products (\ref{prodC}) with the kernel (\ref{Ckernel}) are exponentially suppressed like
$e^{-(z_0 |F''_{ab}|)^2}$. This means that when the relation (\ref{distance}) on the initial wave packets is satisfied,
typicality condition (\ref{wtyp1}) on correlation functions is
fulfilled within a very good level of approximation.

The generalization of the result (\ref{wtyp2}), yielding the
cancellation of $N^{2k-2}n^2$ terms in the variance, cannot be
extended without further assumptions to the case of expanding
modes in the far-field regime. In fact, the relation $|\psi_a(z,t)|=|\psi_b(z,t)|$ might not even be approximately
verified in the large-time limit. Condition $\mathcal{J}^{({C}_k)}$, which, together with the results
discussed above, leads to $D_{2k-2,2}^{({C}_k)}=0$, can be
verified only if the low-momentum Fourier components of the two
modes are approximately equal [see Eq.~(\ref{farfield}).] In the
case analyzed in Ref.~\cite{typbec}, in which the correlation
${C}_2(z)$ has been studied for a system with two translated
Gaussian modes, it turned out that $D_{2k,2}^{({C}_2)}\simeq
0$, as expected from the present general discussion.

\section{Conclusions and outlook}
\label{sec-concl}

In this article we have defined and characterized the typicality of a generic observable in a two-mode Bose system. The results enable
one to determine if an observable is typical once the mode wave functions are known, and to identify different regimes in the fluctuations around the typical expectation value, as the dimension of the sampled subspace varies. The observable analyzed in the
last section are experimentally accessible, and can provide a test for typicality.

The identification of typicality criteria for observables helps one understanding which properties are shared by the vast
majority of states and which ones have instead wide fluctuations.
Remarkably, this distinction is central in determining ``good'' (macroscopic) observables in both classical and quantum statistical mechanics \cite{landau}. The relation between the results obtained in this article and statistical mechanics will be the object of future
research.

It would also be interesting to extend the formalism in order to
include more general cases, like nonuniform samplings or randomly
fluctuating modes. Other possible avenues for future investigation
will be the analysis of the dynamical effects of typicality
\cite{FNPPSY} and the statistical interpretation of recent
experiments on phase randomization in condensates
\cite{Schmied,Schmied2,Schmied3,Schmied4,Schmied5}, as well as the
characterization of the typicality of entanglement in a
Bose-Einstein condensate \cite{benatti}.

\acknowledgments
This work was partially supported by the Italian National Group of Mathematical Physics (GNFM-INdAM), and by PRIN 2010LLKJBX on ``Collective quantum phenomena: from strongly correlated systems to quantum simulators.''

\end{document}